\documentclass[conference]{IEEEtran}

\usepackage[T1]{fontenc}
\usepackage{hhline}
\usepackage{amssymb,amsfonts,amsthm}
\usepackage{amsmath}
\usepackage{epsfig}
\usepackage{color}
\usepackage{enumerate}
\usepackage{ifpdf}
\usepackage{graphicx}
\usepackage[ruled,vlined,linesnumbered]{algorithm2e}
\usepackage[tight,footnotesize]{subfigure}
\usepackage{cite}

\begin{document}

\title{Efficient Communication over Cellular Networks with Network Coding in Emergency Scenarios}

\author{Tan Do-Duy, M. Angeles Vazquez Castro\\
Dept. of Telecommunications and Systems Engineering\\
Autonomous University of Barcelona, Spain\\
Email: \{tan.doduy, angeles.vazquez\}@uab.es}

\maketitle
\begin{abstract}
Emergency communications requires reliability and flexibility for disaster recovery and relief operation. 
Based upon existing commercial portable devices (e.g., smartphones, tablets, laptops), we propose a network architecture that uses cellular networks and WiFi connections to deliver large files in emergency scenarios under the impairments of wireless channel such as packet losses and intermittent connection issues. Network coding (NC) is exploited to improve the delivery probability. We first review the state-of-the-art of NC for emergency communications. Then, we present the proposed network architecture which utilizes multiple radio interfaces of portable devices to support data delivery. A random linear NC scheme is exploited at source to enhance the reliability for content delivery against packet losses. Besides, an analytical model for the successful decoding probability in linear NC is derived. Finally, we evaluate the effectiveness of the proposed architecture with NC in terms of the delivery ratio of content for intermittent connectivity scenarios.

\end{abstract}


\IEEEpeerreviewmaketitle

\section{Introduction}
\label{sec:intro}
Emergency communications in disaster recovery and relief operation requires efficient, robust, and rapid communication networks to deliver data between disaster management headquarter and on-site teams. The efficiency and timeliness of the response is often contingent due to the dynamic and resource limited constraints in the disaster areas \cite{Krishnaswamy.2014}. Therefore, the need of a reliable and robust network infrastructure for communication which ensures available connections to all users in a disaster area is crucial \cite{Cabrera.2013}. 

By popularity of portable devices and cellular networks in everyday communications, such devices become promising candidates to contribute on supporting disaster recovery and relief operation. 
Furthermore, nowadays commercial mobile devices are equipped with advanced signal processing capabilities and multiple radio interfaces such as cellular communication, WiFi, and Bluetooth. They provide a broad range of applications including videos and photos sharing. In addition, more portable devices will be brought to the disaster area by relief workers.

\begin{figure}[htbp]
  \begin{center}
    \epsfig{file=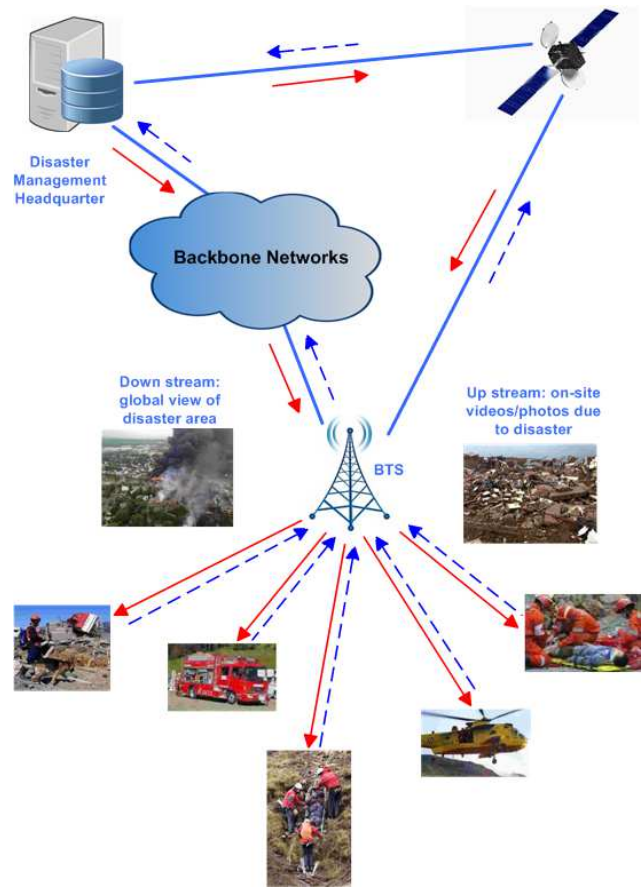,width=0.46\textwidth}
    \caption{Data transfer and distribution scheme for disaster recovery and relief operation. First responders (e.g., on-site relief teams) and the remote control station (e.g., disaster management headquarter or local government) communicate through either backbone networks or satellite communications.}
    \label{fig:model1}
  \end{center}
\end{figure}

A possible scenario is illustrated as in Fig. \ref{fig:model1}. First responders (e.g., on-site relief teams) connect to the remote control station (e.g., disaster management headquarter or local government) through either backbone networks or satellite communications. On the up-stream, first responders in the emergency zone take and share their location and videos/images with the remote control station and each others as well. On the down-stream, the responders request the global view of the emergency area from the remote control station. 
However, due to the effects of wireless channel in the disaster area, the data dissemination may either require retransmission or eventually be disconnected due to random obstacles on the path and packet drops. Therefore, it must be protected against intermittent connectivity issues and unreliable wireless channel. 

In this paper, we focus on the down-stream communication between the remote control station and the responders.
By exploiting NC \cite{Medard.2011} with the existing mobile devices, we propose a novel solution of reliable data dissemination for emergency scenarios, where mobile devices establish simultaneously two network interfaces consisting of the purely cellular communication links to the base station and short range links (e.g., WiFi ad-hoc mode) to neighboring devices within its proximity to ensure that the requested content is available for all responders/relief workers within the disaster area. In this study, a large file is segmented into independent fragments. With NC at source, fragments are linearly combined into network-coded packets to be sent. Each demanded device is responsible for collecting and decoding to recover the entire file from these fragments. 
The benefits of the proposed network architecture with using NC are as follows:

\begin{itemize}
	\item The two radio interfaces guarantee content delivery and connectivity for all devices within the disaster area under effects of wireless channel, i.e. reliable and robust connections for emergency communications.
	\item NC makes efficient use of the limited available bandwidth, especially in case of emergency scenarios.
\end{itemize}


The rest of this paper is organized as follows. 
In Section \ref{sec:survey}, we review the state-of-the-art of NC for emergency communications.
In Section \ref{sec:model}, we derive the proposed network architecture.
In Section \ref{sec:NC}, we present the exploited NC scheme and respective analytical model.
In Section \ref{sec:sim}, we evaluate the effectiveness of our proposed architecture with NC.
Finally, in Section \ref{sec:concl}, we conclude the paper.

\section{A review on Network Coding for Emergency Communications}
\label{sec:survey}
\subsection{Network Coding Overview}
\begin{figure}[htbp]
  \begin{center}
    \epsfig{file=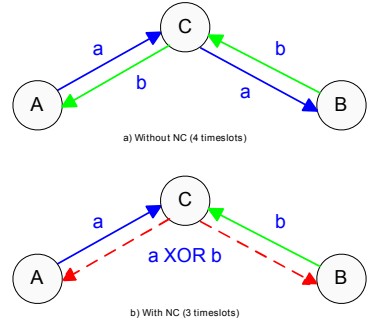,width=0.26\textwidth}
    \caption{An example of NC: a) without NC and b) with NC.}
    \label{fig:xor}
  \end{center}
\end{figure}

NC over the store-and-forward paradigm provides a new solution of enhancing reliability, throughput enhancement, and network design and operation \cite{Pahlevani.2014}. 
For example, a simple three-node model is illustrated in Fig. \ref{fig:xor}, where A and B want to exchange their own packets a and b, respectively. Assume that these nodes are out of range of each other and in time-division access mode, then this communication requires four timeslots including two timeslots for sending the packets to the relay C and two timeslots for relaying the packets. However, with NC, the relay can simply XOR the packets and send the coded packet. Then, both A and B can retrieve the required packet from the other node using their own packets. By this way, the total timeslots for transmission reduce from 4 to 3.

NC can be classified as either inter-session or intra-session \cite{Ostovari.2013}. The former focuses on solving bottleneck problems and reducing the number of transmissions by allowing packets from different sources/flows to be coded together. Therefore, NC decreases the interference between the links in wireless network and increases the overall network throughput. This technique has low computational complexity for coding. However, its drawback is not resilient to packet losses in the system. On the other hand, intra-session NC leads reliability enhancement in wireless networks with smaller number of transmissions than the feedback-based scheme without NC. However, it requires higher computational complexity than inter-session scheme. This approach exploits the link diversity by combining different packets from the same source/flow. Intra-session NC usually relies on random linear NC (RLNC) to encode and decode packets in a group with coefficients chosen from a finite field. This field size determines the probability that the destination can obtain linearly independent combinations and therefore obtain innovative information to recover the original packets successfully.

In general, NC has recently emerged as a new approach for improving network performance in terms of throughput and reliability, especially in wireless networks by the uncertainty of wireless medium. 
This section briefly presents a literature review of the state-of-the-art of NC for emergency communications, including general emergency cases, emergency cases in vehicular ad-hoc networks (VANETs), and large potential of NC applications for commercial mobile devices. 

\subsection{Network Coding for General Emergency Cases}
Emergency communications should be reliable and flexible for disaster aid and relief operation \cite{Lee.2010}. 
Reliability, availability and robustness have been considered as fundamental requirements for broadband communications and networks during disaster and emergency times. NC is a promising solution to enhance to reliability and robustness for data transmission.

Joy et al. \cite{Joy.2013} presented an implementation of network infrastructure with NC to deliver large files from a source to a destination with the help by surrounding nodes, e.g. a real-time video from a cellphone to a helicopter. Intra-session NC is applied at source node, then surrounding nodes forward overhearing packets. At the destination, decoding procedure is done to recover the original file. The NC helps to improve the numbers of files delivered compared to fragmentation in scenarios of packet loss or disruptions. The advantages come from spatial diversity by surrounding nodes. Various nodes may repeat different pieces of a file due to link disruption by channel condition or busy relays. Besides, the relay may recover the original files from different pieces to use before forwarding. 
Nevertheless, this paper only exploited WiFi ad-hoc mode on the Android phones and laptops in short-range communications.

In \cite{Altamimi.2014}, NC is employed to improve the delivery probability in an intermittently connected network (ICN), which utilizes mobile networks with cooperation between nodes to create message replication. Main targets are to maximize the delivery ratio and minimize the overhead ratio. The authors showed an explicit expression for the delivery probability of random linear NC in comparison to the normal replication. This work requires overhearing ability for each device in mobile networks. 

Besides, Nguyen et al. \cite{Hung.2014} designed a novel NC aided MIMO scheme for combating the deleterious effects of both the shadow fading and Rayleigh fading in hostile wireless channels. The proposed model leads to ambulance-and-emergency communications. A powerful space-time code is proposed for providing a near-capacity performance in fast fading environments. NC is herein used to obtain a further spatial diversity gain for combating slow fading effects by obstacles. 
In \cite{Alejandra.2014}, a novel perceptual semantics for multimedia communications is proposed to enhance situation awareness in human-analysis-driven processes as in emergency operations. However, this study mainly focuses on application layer optimization with adaptive NC at network layer, which assumes a network architecture as a FIFO with finite queue.
In addition, authors in \cite{Subramanian.2010} considered an intermittently-connected mobile network consisting of N relays, 1 source and M destinations. NC is utilized to enhance the transmission capacity limited by disruptive connectivity. Each relay makes random linear combination for incoming packets over $GF(q=2^{F})$ before sending to the others. Queuing-theoretic is derived to analyze the steady-state throughput performance of the network-coded scheme.


For VANETs, beacon information plays an important role in vehicle applications such as predicting the position of neighbor vehicles to avoid any emergency situation \cite{Sahu.2014a}. However, beacon overhead and congestion may cause low message reception, which affects emergency messages and other control messages. The work in \cite{Sahu.2014a} considered packet level NC for controlling beacon overhead. An intermediate vehicle considering as a relay will perform $XOR$ operation for incoming packets from other two vehicles, i.e. $C$= $A$ $XOR$ $B$. Then, the combined packet is forwarded back to the two sources. By this way, channel contention caused by beacon overhead is reduced. Besides, the authors in \cite{Sahu.2014b} also utilized $XOR$ operation for incoming packets from other two vehicles as in \cite{Sahu.2014a}. However, the target is to cancel interferences due to inter-street beacon communications by adaptive transmission control. 

\subsection{Network Coding on Commercial Mobile Devices}
By huge number of devices, implementing NC on commercially mobile devices via cellular networks and WiFi connection opens a promising approach for real-time applications such as emergency communications, where real-time videos/photos can be taken and sent in a timely matter.

In \cite{Peng.2012}, architectures of network-coded have been presented for the next generation IMT-Advanced systems. The key point is that the collaboration by the intermediate nodes can enhance network performance significantly. However, the existing 3G/4G cellular networks have not supported for cooperative relay schemes and decoding procedure at the base station as well. Moreover, respective scheduling schemes are also required. Therefore, this work is a first look at the possible scenarios to the design of cooperative NC in futuristic wireless communication systems.

Pedersen et al. \cite{Pedersen.2008} implemented a simple scheme of NC at intermediate cellphones. By combining both 3G/4G links and WiFi connection on cellphones, the overall network bandwidth may be reduced by 50 percents. Even though the model is simple, this work opens an new approach to apply NC for existing cellular network devices, which is not feasible by \cite{Peng.2012}.
In addition, in \cite{Pahlevani.2014}, many NC schemes for data sharing have been implemented on commercial cellphones. The authors considered both 3G/4G and WiFi connection between devices. 

In general, realistic applications of NC for emergency communications over cellular communication systems have not been considered significantly due to the limitations such as cooperative relay schemes, decoding procedure, and scheduling schemes at the base station.
However, if both 3G/4G and WiFi are flexibly combined, NC can bring enhancement on network utilization efficiency and reliability over the hostile channel conditions.

\section{Description of the Proposed Network Architecture}
\label{sec:model}
\begin{figure}[htbp]
  \begin{center}
    \epsfig{file=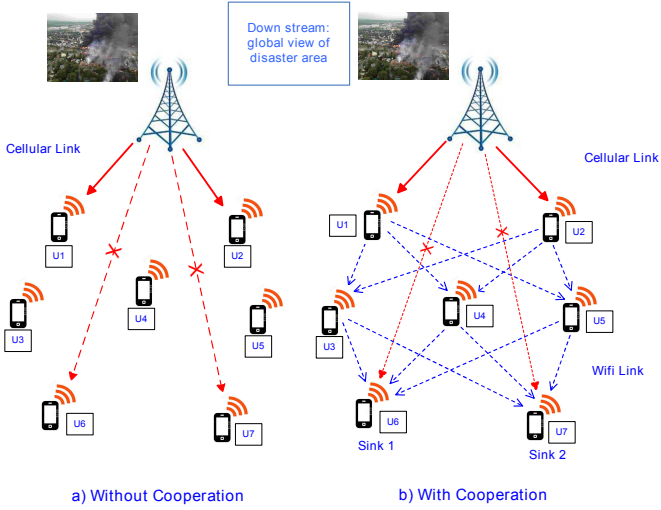,width=0.48\textwidth}
    \caption{Network model: a) Direct cellular links between BTS and users without cooperation, b) Cooperation between cellular links and WiFi links for data download and distribution. The red dotted lines denote failure or intermittent connections over cellular links; however, the proposed transmission strategy may provide data dissemination by other connections over WiFi links.}
    \label{fig:model2}
  \end{center}
\end{figure}

Assume that the purely cellular network infrastructure is still functional under the effects of disasters, mobile devices will be helpful in the efficient distribution of rescue and relief to disaster area. However, for communication in this environment, end-to-end connectivity cannot be always guaranteed to all users in the field, where the construction of a continuous end-to-end path between source and destination is difficult or impossible.
In case of large-scale disasters such as flood and cyclone, cellular network infrastructure may immediately become non-functional due to system damage. The proposed network scheme in Fig. \ref{fig:model2} can then be adapted by replacing the base station with a portable wireless station which is equipped with satellite communications and short range radio links.

In Fig. \ref{fig:model2} (b), we introduce a novel network architecture based on existing mobile devices in cellular networks, where content delivery from the disaster management headquarter is guaranteed to all users in intermittent connectivity scenarios. Moreover, overall network resource on cellular communications may be saved by cooperation between devices over cellular links and WiFi links for data download and distribution.

Assume that relief workers with mobile devices in disaster area request the same multimedia content from the headquarter, e.g. global view of the disaster area. We can consider these users as a multicast group as illustrated in Fig. \ref{fig:model1}. The network model consists of two groups. The first group of users is connected directly to the cellular base station through cellular links. However, due to the channel effects or obstacles which cause either disconnection or intermittent connectivity, the other users cannot work well with cellular links. Thus the second group, which considers the first group as indirect sources to relay the requested multimedia content, forms an ad-hoc network based on WiFi links. This model allows data dissemination from the headquarter to all on-site relief workers in emergency scenarios, especially in case of intermittent connectivity due to obstacles or difficult terrain at the disaster areas. 

\section{Random Linear Network Coding Scheme}
\label{sec:NC}

\subsection{Network Coding Scheme}
\label{sec:des}

\begin{figure}[htbp]
  \begin{center}
    \epsfig{file=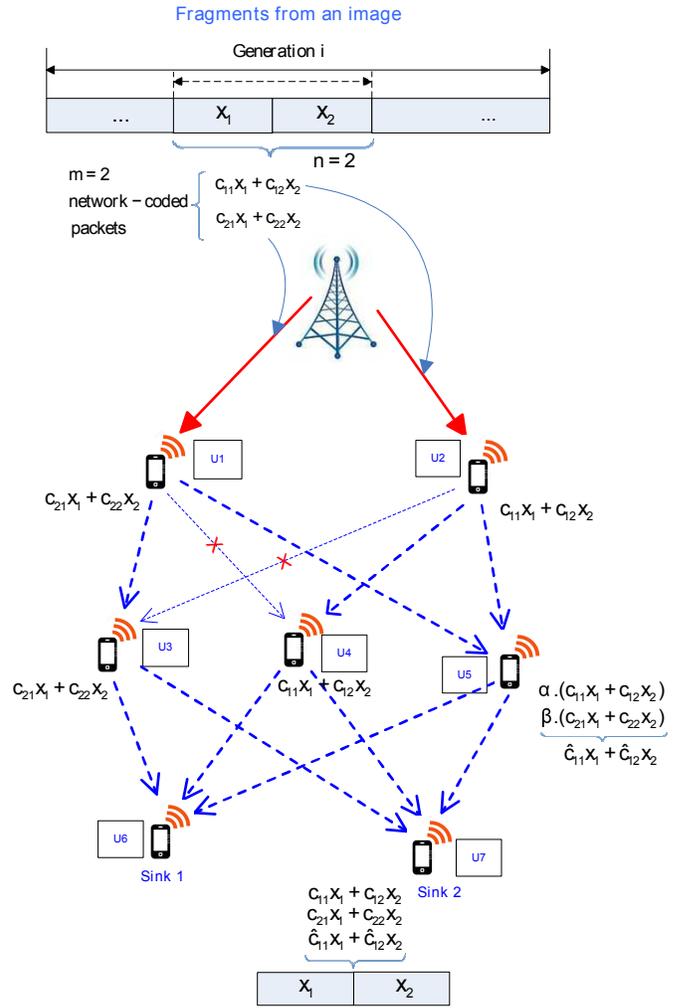,width=0.48\textwidth}
    \caption{Network-coded traffic with random linear NC at source. For example, in case of $n=2,m=2$, each coded packet is transmitted to a relay in the first group via a cellular link. Then, these packets are forwarded toward the next hop via WiFi connections. Intermediate nodes in the second relay group can recombine multiple received packets from the same generation to increase linear independence for received packets at the sinks. Even if erasure events occur on one of incoming links of the relays, delivery performance is still guaranteed at the destination.}
    \label{fig:model3}
  \end{center}
\end{figure}

We consider the proposed network architecture in Fig. \ref{fig:model2}(b), where a multicast group requests an image of the disaster area sent from the management headquarter (source node). For simplicity, we assume that the base station plays the same role as the source node. 
To improve the reliability for content delivery against packet losses, the source exploits a random linear NC scheme to the file before transmission as illustrated in Fig. \ref{fig:model3}.
First, the file is segmented into a set of fragments or blocks. The source then linearly combines $n$ fragments $x_{i}$ $(1\leq i\leq n)$ to generate $m$ $(m \geq n)$ linear combinations $y_{j}$ $(1\leq j\leq m)$ as follows
\begin{eqnarray}
{y_{j}}	= \sum_{\substack{1\leq i\leq n}} c_{ji}x_{i},
 \label{eq:rlnc1}
\end{eqnarray}
where $c_{ji}$ is a coefficient randomly generated from a Galois field of size $q=2^{f}$ ($c_{ji}\in \mathbb{F}_{q} \backslash \left\{0\right\}$), e.g. $f=8$. A set of $m$ linear combinations produced from $n$ fragments is assigned the same generation number.

In general, let $X\in \mathbb{F}^{n \times 1}_{q}$ denote the $n\times 1$ vectors of source fragments and $C\in \mathbb{F}^{m \times n}_{q}$ with rank $n$ denote the $m \times n$ coefficient matrix. Then, the $m \times 1$ vectors of the transmitted network-coded packets $Y\in \mathbb{F}^{m \times 1}_{q}$ are given by
\begin{eqnarray}
Y	= C \cdot X.
 \label{eq:rlnc2}
\end{eqnarray}

At intermediate nodes, multiple packet combinations from the same generation are linearly recombined to generate at the output, where each element is a randomly coefficient from the same finite field $GF(2^{f})$.
At the sink, if it received at least $n$ linearly independent encoded messages from the source node, the network-coded packets are decoded by 
\begin{eqnarray}
\hat{X} = \hat{C}^{-1} \cdot \hat{Y},
 \label{eq:rlnc3}
\end{eqnarray}
where $\hat{C}^{-1} \in \mathbb{F}^{m^{'} \times n}_{q}$ ($m^{'} \geq n$) are the $m^{'} \times n$ coefficient matrix relevant to the received packet combinations $\hat{Y}$.


\subsection{Decoding Probability Analysis}
\label{sec:analysis}
\begin{figure}[htbp]
  \begin{center}
    \epsfig{file=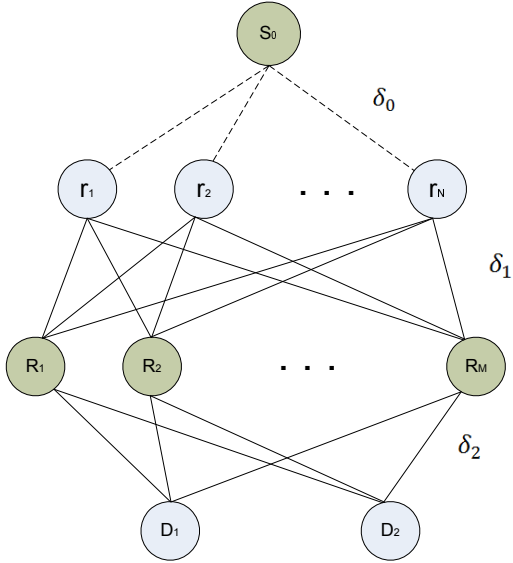,width=0.38\textwidth}
    \caption{A linear NC scheme at source $S$ and $N+M$ relays.}
    \label{fig:model4}
  \end{center}
\end{figure}

In this section, we investigate the successful decoding probability in random linear NC at source with erasure channels. 
For a generalized analysis of the proposed network architecture in Fig. \ref{fig:model3}, we consider a relay network composed of 1 traffic source $S$, $N$ and $M$ relay nodes at the first and the second hops, respectively, and a number of destinations, as denoted in Fig. \ref{fig:model4}. All links are assumed to be independent channels. Random linear NC scheme follows the description in Sec. \ref{sec:des}.
The source generates $m$ linear combinations from each of $n$ original fragments via different links to the first hop neighbors $r_{i}$ $(1\leq i\leq N)$. The network-coded packets forwarded by the first group are then re-encoded at the second hop nodes $R_{j}$ $(1\leq j\leq M)$ before reaching the destinations.

The effectiveness of random linear NC depends on the availability of at least $n$ linearly independent encoded packets for each generation at the destination to recover the original data. This condition relates to the impacts of erasure rate and NC design. 
Therefore, we derive the decoding probability at the destination as a function of the packet loss rate, coding design parameters $(n,m)$, and the number of relays $(N,M)$.

Let $\delta_{0}$ be the erasure rate for the links between $S$ and $r_{i}$, $\delta_{1}$ be the erasure rate for the links between $r_{i}$ and $R_{j}$, and $\delta_{2}$ be the erasure rate for the link between $R_{j}$ and the destinations. 
On the erasure links between $R_{j}$ and the destination, we let $\overline{\epsilon}$ denote the successful reception event with $Pr\left\{\overline{\epsilon}\right\}=1-\delta_{2}$ and $\epsilon$ denote the occurrence of an erasure event with $Pr\left\{\epsilon\right\}=\delta_{2}$.

We consider the extraction of each element $\hat{c}_{ji}$ in coefficient matrix $\hat{C}$ at the destination under the effects of erasure channels $\delta_{0}$, $\delta_{1}$, and $\delta_{2}$.
At a specific relay $R_{j}$ (e.g., $R_{1}$), it can be observed that in case of without packet loss on the link $R_{1}$ to $D$, the random element is equal to zero only if packet losses have occurred on either all $N$ links from the $r_{i}$ to $R_{1}$ or all $N$ links from the source to $R_{j}$. On the other hand, the element is a random value from a finite field excluding zero. Therefore, the conditional probabilities are respectively defined as 
\begin{eqnarray}
Pr\left\{\hat{c}_{ji}= 0 | \overline{\epsilon}\right\}	=	\left[ \left(1-\delta_{0}\right)\delta_{1} \right]^{N} + \left(\delta_{0}\right)^{N}	=	\psi,
 \label{eq:a1}
\end{eqnarray}
\begin{eqnarray}
Pr\left\{\hat{c}_{ji}= \theta | \overline{\epsilon}\right\}	=	\left(1-\psi \right)/\left(q-1 \right) (\theta \neq 0).
 \label{eq:a2}
\end{eqnarray}

We adapted the analytical model in \cite{Seong.2014}, which is based on rank of coefficient matrix to derive the bound for the decoding probability. 
The delivery failure probability $P_{fail}$ is defined as 
\begin{eqnarray}
P_{fail} 	&	:=		&	Pr\left\{rank(\hat{C}) < n	\right\}																										\nonumber \\
					&	=			&	Pr\left\{\exists v: \hat{C}v	=	0^{T}	\right\}																					\nonumber \\
					&	\leq	&	\sum_{\substack{v \in \mathbb{F}^{n}_{q} \backslash \left\{0^{T}\right\}}}	Pr\left\{\hat{C}^{-1}v	=	0^{T}	\right\},
 \label{eq:a3}
\end{eqnarray}

where $v$ is a nonzero vector with $n$ elements and $0^{T}$ denotes a zero vector.

Follow the same approach in Theorem $1$ in \cite{Seong.2014}, we obtain
\begin{eqnarray}
P_{decode}	=	1	-	P_{fail},
 \label{eq:a4}
\end{eqnarray}
where
\begin{multline}
P_{fail}	\leq	\frac{1}{q-1}	\sum_{\substack{1\leq i \leq n}}	\dbinom{n}{i}(q-1)^{i} \bigg\{\delta_{2} + \left(1-\delta_{2}\right)	\\
								\times \bigg[ q^{-1} + \left(1-q^{-1} \right) \left(1-\frac{1-\psi}{1-q^{-1}}\right)^{i} \bigg] \bigg\}^{M\cdot\frac{m}{N}}.
 \label{eq:a5}
\end{multline}


For the sake of comparison, we also derive the decoding probability in inter-NC scheme at the second hop relays based on the same network model as in Fig. \ref{fig:model4}. Where the source generates $N$ purely fragments without combination. Each of the $M$ intermediate nodes $R_{j}$ encodes the received messages using random linear NC before forwarding. The decoding probability of the $N$ fragments is given by
\begin{multline}
P^{'}_{decode}	\geq (1 - \delta_{0})^{N} \times \bigg\{ 1 -	\frac{1}{q-1}	\sum_{\substack{1\leq i \leq N}}	\dbinom{N}{i}(q-1)^{i} \\
\times \bigg\{\delta_{2} + \left(1-\delta_{2}\right)
\bigg[ q^{-1} + \left(1-q^{-1} \right) \left(1-\frac{1-\delta_{1}}{1-q^{-1}}\right)^{i} \bigg] \bigg\}^{M}\bigg\}.
 \label{eq:a6}
\end{multline}

\begin{figure}[htbp]
  \begin{center}
    \epsfig{file=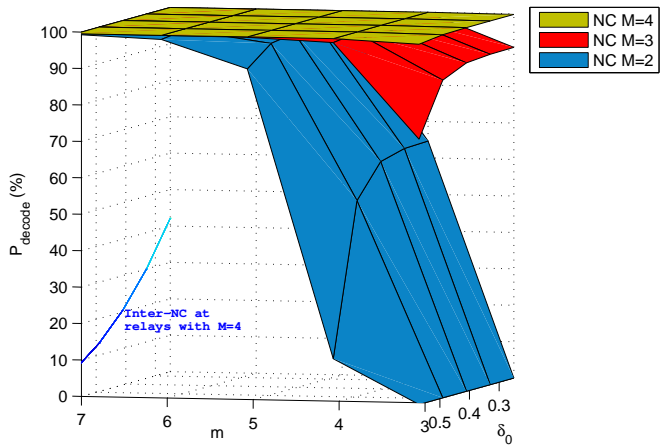,width=0.55\textwidth}
    \caption{Theoretically decoding probability of NC at source according to $q=256$, $n=3$, $N=3$, $\delta_{1}=0.01$, $\delta_{2}=0.01$.}
    \label{fig:theory3D}
  \end{center}
\end{figure}

Fig. \ref{fig:theory3D} plots the bound for the decoding probability with respect to the number of relays $M$ at the second hop as a function of erasure rate $\delta_{0}$ on the links between the source and the first hop relays and coding design parameter $m$. Each surface corresponds to a different value of $M$.
In addition, by the solid blue line, we compare theoretical performance of random NC at source and inter-NC with respect to various values of erasure rate $\delta_{0}$. We can observe that inter-NC at intermediate relays with $M=4$ is dramatically affected by erasure channels between the source and the first hop relays.
On the other hand, random linear NC at source with link diversity can potentially provide better delivery performance by flexibly changing NC parameters $m$ at the source node. In particular, in case of $M=4$, the decoding probability at destinations approximately reaches a maximum value of $100\%$ according to erasure rates in the range between $0.2$ and $0.55$. This is because of that the redundancy of intermediate relays $R_{j}$ at the second hop ($M=4$ versus $N=3$) combining with link diversity by packet combination at the source increases opportunity for the network-coded packets to reach the destinations. On the other case, assume that the available relays are reduced to $M=2$ while coding parameters at the source are designed to be $n=3$ and $m=4$. Then, we see a significant degradation of the decoding probability if compared with the other cases. However, the problem can be improved by increasing redundant combinations at the source, i.e. $m-n$ packets. For example, with $m=5$, the decoding ratio obtains a value of $90\%$ at $\delta_{0}=0.5$.

The above observation reveals a practical aspect of NC at source, called \textit{Geo-Network coding}, which takes into account the geographical information of relays in the emergency area to network coding design so as to enhance robust and reliable transmission and delivery in situation awareness scenarios. Assume that the source has access to coverage maps and locations of relays, Geo-Network coding then simply mean to select the $M$ appropriate relays depending on the signal strength at their locations. Even if the number of relays $R_{j}$ is not available to provide a required performance, the source can increase the network performance itself by the $m-n$ redundant combinations.

\section{Simulation Results}
\label{sec:sim}
Assume that intermittent connectivity due to obstacles or difficult terrain in the disaster areas prevents direct transmission from the base station to the demanded users using cellular links. In this case, our architecture is a feasible approach to support service to the users. We take benefits of NC to enhance the reliability for multicast data.
In this section, we evaluate the performance of the proposed network architecture in multicast data delivery with using NC and without using NC in terms of average packet delivery ratio (PDR) at the sinks. 
The simulated network scheme is illustrated as in Fig. \ref{fig:model3}, where the source transmits multicast data using different channels via cellular links to some intermediate nodes in the group. These packets are then forwarded hop-by-hop to their neighboring nodes via WiFi links before reaching the sinks. 

The objective of simulation is to send a multicast file from a source to 2 sinks through intermediate relays, where 2000 packets are transmitted across the simulated network. The Galois Field size for NC is $2^8$, which is sufficiently large enough for practical applications. Link packet erasure patterns at different erasure rates are generated using Gilbert Elliot model. 
We compare the delivery performance of purely fragmentation without NC and fragmentation with NC at source versus per link packet erasure probability.

\begin{figure}[htbp]
  \begin{center}
    \epsfig{file=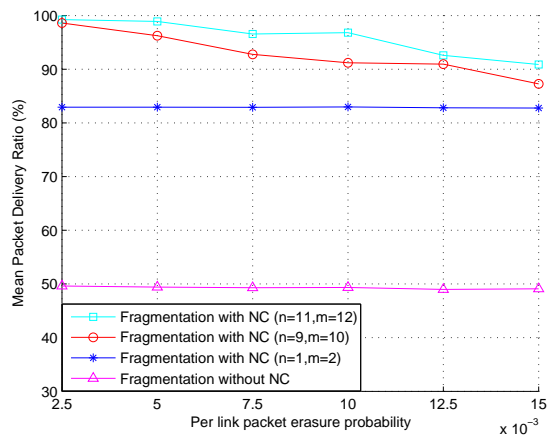,width=0.45\textwidth}
    \caption{Mean delivery ratio according to different transmission strategies versus per link packet erasure probability.}
    \label{fig:LossRate}
  \end{center}
\end{figure}

Initial simulations are performed to evaluate the average PDR at multiple sinks versus varying link erasure probabilities. Fig. \ref{fig:LossRate} compares the performance of fragmentation without NC and fragmentation with NC assigned different values of $(n,m)$, i.e. the $m$ linear combinations for each $n$ source fragments. 
We can observe that purely fragmentation with only store-and-forward presents very low PDR due to erasure channels and packet drop at intermediate nodes. 
On the other hand, fragmentation schemes with NC generally outperforms the purely fragmentation regardless of the effect of erasure channel. This is because of NC's ability to recover from losses, i.e. reliability over erasure channels, by exploiting link diversity and re-encoding packets with the same generation at intermediate nodes to reduce congestion, which increases probability of successfully transmitting a coded packet from the source to the sinks. In particular, the case of NC with $n=11, m=12$ obtains the highest performance by its higher link diversity than the other schemes although its redundancy ratio is a bit smaller than the case of NC with $n=9, m=10$ ($1/11$ and $1/9$, respectively). Whereas, in case of $n=1, m=2$, because of small link diversity, its PDR is the smallest one if compared to the other NC cases. However, it still obtains a significant improvement on network performance with respect to various packet loss rates.  
In general, the revealed simulation results show that hybrid network architecture with NC can considerably improve multicast data delivery in scenarios with bandwidth-constrained and severe disruptions such as intermittent connectivity and erasure channels.

\begin{figure}[htbp]
  \begin{center}
    \epsfig{file=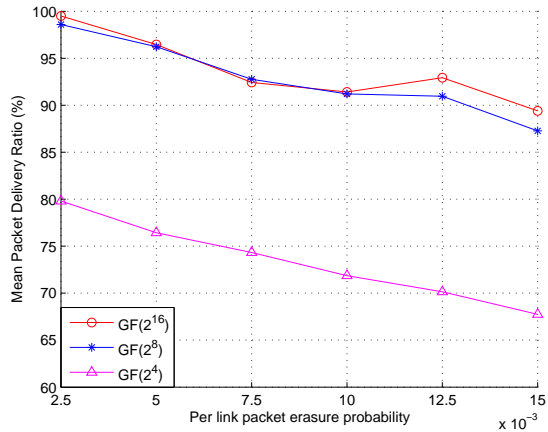,width=0.45\textwidth}
    \caption{Mean delivery ratio of fragmentation with NC in case $n=9$, $m=10$ according to different field sizes $GF(2^k)$ versus per link packet erasure probability.}
    \label{fig:GFsize}
  \end{center}
\end{figure}

The subsequent results show the impacts of Galois Field size on network performance. Fig. \ref{fig:GFsize} denotes mean delivery ratio versus various link erasure probabilities according to different field sizes. The larger field size conduces the higher delivery ratio at the destination. 
The reason is that the larger the field size, the larger the probability that the sinks receive $n$ linearly independent encoded packets from the source, i.e. the larger the probability of successful decoding a network-coded packet generation.


\section{Conclusion}
\label{sec:concl}
In this paper, we proposed a novel network architecture based on the existing cellular networks and commercial mobile devices for intermittent connectivity scenarios. By exploiting random linear NC, we can guarantee robust and reliable content delivery and connectivity for all devices within the disaster area under the effects of wireless channels and obstacles. Simulation results show that in the proposed network architecture, fragmentation with NC at source significantly outperforms the purely fragmentation scheme in terms of the delivery probability. 
The impacts of finite field size in network performance is also evaluated.
This work is only a first step toward a full consideration of novel network architectures for emergency scenarios. In the future, we plan to investigate real implementation and performance evaluation of the proposed architecture on the existing commercial cellphones.

\section*{Acknowledgments}

\bibliographystyle{IEEETran}

\bibliography{emerge}

\end{document}